\documentclass[pdflatex,sn-mathphys-num]{sn-jnl}

\usepackage{graphicx}%
\usepackage{multirow}%
\usepackage{amsmath,amssymb,amsfonts}%
\usepackage{amsthm}%
\usepackage{mathrsfs}%
\usepackage[title]{appendix}%
\usepackage{xcolor}%
\usepackage{textcomp}%
\usepackage{manyfoot}%
\usepackage{booktabs}%
\usepackage{algorithm}%
\usepackage{algorithmicx}%
\usepackage{algpseudocode}%
\usepackage{listings}%
\newcommand{\PTO}{PbTiO$_3$}
\usepackage{gensymb}
\usepackage{siunitx}
\newcommand{\note}[1]{{\color{black}{#1}}}

\begin{document}

\title[Disentangling the Discrepancy Between Theoretical and Experimental Curie Temperatures in Ferroelectric PbTiO$_3$]{Disentangling the Discrepancy Between Theoretical and Experimental Curie Temperatures in Ferroelectric PbTiO$_3$}

\author[1]{\fnm{Denan} \sur{Li}}\email{lidenan@westlake.edu.cn}

\author[1]{\fnm{Christian S.} \sur{Ahart}}\email{chris\textunderscore ahart@westlake.edu.cn}

\author*[1,2]{\fnm{Shi} \sur{Liu}}\email{liushi@westlake.edu.cn}

\affil[1]{\orgdiv{Department of Physics, School of Science}, \orgname{Westlake University}, \orgaddress{\city{Hangzhou}, \postcode{310030}, \state{Zhejiang}, \country{China}}}
\affil[2]{\orgdiv{Institute of Natural Sciences}, \orgname{Westlake Institute for Advanced Study}, \orgaddress{\city{Hangzhou}, \postcode{310024}, \state{Zhejiang}, \country{China}}}


\abstract{Predicting ferroelectric Curie temperatures ($T_c$) from first principles often underestimates experimental values. We investigated this discrepancy in PbTiO$_3$ using constant-pressure \textit{ab initio} molecular dynamics and machine learning force fields (MLFFs). Our results indicate the underestimation stems primarily from exchange-correlation functional limitations rather than MLFF inaccuracies. We reveal that short-range MLFFs match experiments through error cancellation, whereas including physical long-range interactions lowers $T_c$. Consequently, accurate predictions require large simulation cells, explicit long-range interactions, and improved functionals.}


\maketitle

\section{Introduction}\label{sec1}

Ferroelectric materials are characterized by a spontaneous electric polarization that is reversible under an external electric field. The Curie temperature ($T_c$), the critical point at which the material undergoes a phase transition from a ferroelectric to a paraelectric phase, serves as both a fundamental hallmark of ferroelectricity and a practical constraint in device design~\cite{Lines77}. Its magnitude indicates the strength of dipole-dipole interactions and the energy scale of symmetry-breaking transitions~\cite{Abrahams68p551}. Technologically, $T_c$ defines the operational temperature range of ferroelectric-based devices, as properties such as dielectric susceptibility change significantly near the transition~\cite{Pan19p578,Liu25p211}. 

Despite its significance, the direct prediction of $T_c$ from composition using first-principles density functional theory (DFT) remains challenging. As a zero-temperature ground-state theory, standard DFT must be extended to account for finite-temperature effects, such as anharmonic lattice dynamics and domain evolution, in order to accurately describe phase transitions~\cite{Zhong94p1861,ghosez22p325}. While \textit{ab initio} molecular dynamics (AIMD) can, in principle, capture these effects, its high computational cost poses a significant challenge for simulating the large supercells and long time scales required to obtain sufficient configurations for reliable statistical analysis~\cite{Srinivasan03p168,Fang15p024104}.
Previous theoretical studies have often sought to correlate experimentally measured $T_c$ with DFT-derived descriptors such as cation displacements and polarization ($P$)~\cite{Balachandran16p144111,Grinberg07p37603}. For example, Grinberg and Rappe showed that in Bi($B_{1/2}^{2+}$$B_{1/2}^{4+}$)O$_3$–PbTiO$_3$ ($B^{2+}$=Mg, Zn; $B^{4+}$=Zr, Ti) solid solutions, $T_c \propto P^2$, with nonmonotonic $T_c$ trends arising from composition-dependent cation displacements~\cite{Grinberg07p37603}. However, it remains uncertain whether such empirical relationships are broadly applicable across different types of solid solutions.

Machine learning-based force fields (MLFFs) provide a promising way to combine the accuracy of DFT with the speed of classical force fields~\cite{Zhang18p143001,gpumd4.0}. These models make it possible to simulate temperature-driven phase transitions with quantum-level accuracy but at much lower computational cost. However, when applied to well-known ferroelectrics like PbTiO$_3$ (PTO), the results have been inconsistent: the predicted transition temperatures ($T_c$) are often much lower than the experimental value of $\approx760$ K~\cite{Wu23p144102,Wu23p180104,Shi24p174104}.
Such underestimation is not unique to MLFFs. For example, two conventional force fields, the bond-valence model and the shell model, both of which use atomic charges to capture long-range Coulomb interactions, predict transition temperatures of $\approx400$ K~\cite{Liu13p104102} and $\approx580$ K~\cite{Gindele15p17784}, respectively.
This persistent underestimation raises a fundamental question: What is the primary origin of the error in theoretical $T_c$ predictions from MD simulations? Does it stem from approximations in the model potential (fitting error), or does it reflect the intrinsic limit of the exchange-correlation functional used to generate the training data?

In this work, we disentangle the contributing factors through a rigorous benchmark study of bulk PbTiO$_3$. By performing the largest constant-temperature, constant-pressure ($NPT$) AIMD simulations of the ferroelectric transition in PbTiO$_3$ to date and directly comparing them with predictions from classical MD simulations using MLFFs, we show that the $T_c$ underestimation stems primarily from the intrinsic limitations of the exchange-correlation functional, rather than from inaccuracies in the ML model itself. 
Additionally, we find that a simple AIMD-based sampling method performs well for bulk PbTiO$_3$, yielding an accurate potential for simulating the ferroelectric-to-paraelectric phase transition. Furthermore, we uncover an intriguing interplay between finite-size effects and the locality of descriptors of MLFFs: short-range models exhibit increasing errors with system size, which are effectively mitigated by explicitly incorporating long-range interactions. Interestingly, the short-range model yields a $T_c$ value more closely aligned with experimental measurements than the more sophisticated long-range model, suggesting a degree of error cancellation at play.
These findings, along with the high-quality dataset generated in this study, provide valuable guidance for the development of reliable MLFFs capable of accurately capturing highly anharmonic phenomena in complex materials.

\section{Computational Methods}\label{sec2}
\subsection{DFT and AIMD}
All DFT calculations and AIMD simulations were performed using the CP2K/Quickstep package~\cite{cp2kjcp2020}. The core electrons were described by Goedcker-Teter-Hutter (GTH) pseudopotentials~\cite{Goedecker96p1702,Hartwigsen98p3641}, while the valence electrons were treated using double-$\zeta$ valence polarized basis sets. Temperature was controlled via the canonical sampling through velocity rescaling (CSVR) thermostat with a time constant of 200 fs~\cite{Giovanni07p014101}. Pressure was maintained at 1 bar using the Martyna-Tuckerman-Tobias-Klein (MTTK) barostat~\cite{Glenn94p4177}. The integration time step was set to 1 fs. 

For the PbTiO$_3$ system, the Perdew-Burke-Ernzerhof functional modified for solids
(PBEsol) was employed~\cite{Perdew08p136406}, as it is known to provide a more accurate description of structural properties compared to PBE, which significantly overestimates the tetragonality of the ferroelectric phase~\cite{Wu06p235116}. All $NPT$ AIMD simulations were performed using a $4\times4\times4$ supercell containing 320 atoms, with a plane-wave energy cutoff of 1000 Ry. The valence electron configurations were treated as Pb ($6s^2 6p^2$), Ti ($3s^2 3p^6 4s^2 3d^2$), and O ($2s^2 2p^4$). The system was first pre-equilibrated at 300 K using classical molecular dynamics with an MLFF (see details below). Starting from this equilibrated structure, AIMD simulations followed a stepwise heating protocol: the system was equilibrated at each target temperature for 10 ps, followed by an instantaneous temperature increase of 50 K. 
To maintain consistent plane-wave basis quality during volume fluctuations, the reference cell dimensions were fixed at $17\times17\times17$~\AA, slightly larger than the equilibrium lattice constant. For each temperature near $T_c$, a production trajectory of at least 20 ps was used to ensure sufficient statistical sampling for property calculations. \note{The sampling-time convergence of thermodynamic averages and the autocorrelation of the polarization near $T_c$ were further examined using MLFF trajectories (see Figs.~S4 and S5).}
The local polarization of unit cell $m$ at time $t$ was calculated as
\begin{equation*}
\mathbf{P}^{m}(t)=
\frac{1}{V_{\mathrm{uc}}}
\left[
\frac{1}{8} Z_{\mathrm{Pb}}^{*} \sum_{k=1}^{8} \mathbf{r}_{\mathrm{Pb},k}^{m}(t) + Z_{\mathrm{Ti}}^{*} \mathbf{r}_{\mathrm{Ti}}^{m}(t)
+ \frac{1}{2} Z_{\mathrm{O}}^{*} \sum_{k=1}^{6} \mathbf{r}_{\mathrm{O},k}^{m}(t)
\right],
\end{equation*}
where $V_{\mathrm{uc}}$ is the volume of the unit cell; $Z_{\mathrm{Pb}}^{*}$, $Z_{\mathrm{Ti}}^{*}$, and $Z_{\mathrm{O}}^{*}$ are the Born effective charges of Pb, Ti, and O atoms, respectively; and $\mathbf{r}_{\mathrm{Pb},k}^{m}(t)$, $\mathbf{r}_{\mathrm{Ti}}^{m}(t)$, and $\mathbf{r}_{\mathrm{O},k}^{m}(t)$ are the atomic coordinates in unit cell $m$ at time $t$.

\subsection{Machine Learning Force Fields}
In this work, we mainly focus on the performance of the deep potential (DP) model, an MLFF that employs deep neural networks to map local atomic environments to atomic energies, whose sum yields the total energy of the system. The theoretical foundation and implementation details of the DP method have been extensively documented in our previous work; thus, we refer readers to Ref.~\cite{Wu21p024108} for a comprehensive description and omit redundant details here.

The training datasets employed in this work were from our previous study~\cite{Wu23p144102}. To ensure consistency with the present computational setup, all configurations in these datasets were re-evaluated using CP2K, employing the same DFT parameters described in the previous section to recompute the potential energies, atomic forces, and virial tensors. The resulting DP model demonstrates excellent agreement with the reference DFT data, achieving root mean square errors (RMSEs) of 1.74 meV/atom for energy, 83.52 meV/Å for atomic forces, and 11.03 meV/atom for virial tensors.

Neuroevolution potential (NEP) is another popular MLFF that utilizes the separable natural evolution strategy (SNES) to simultaneously optimize both the descriptor and the network weights. Similarly, its descriptor maps local atomic environments using a set of radial and angular components, expanded onto a basis of Chebyshev polynomials, where the basis coefficients are optimized simultaneously with the neural network weights~\cite{gpumd4.0}. Owing to its simple architecture and an efficient implementation relying on native CUDA kernels without external dependencies, NEP achieves exceptional computational performance, making it one of the fastest machine learning potentials currently available~\cite{liang25parxiv}. Recently, the qNEP method was introduced to incorporate long-range electrostatics by augmenting the short-range neural network with a latent charge model~\cite{Fan2026p4787}. In this framework, the network infers atomic partial charges to explicitly calculate the long-range Coulombic energy, enabling the treatment of electrostatic effects with minimal computational overhead and without requiring explicit charge labels~\cite{cheng25p1,kim25p24,Song2024p2088}. \note{While direct Ewald summation can become computationally demanding for large systems, qNEP employs the particle-particle particle-mesh technique to accelerate the evaluation of long-range electrostatics, enabling efficient large-scale simulations with explicit Coulomb interactions.}
To investigate finite-size effects and long-range corrections on $T_c$ predictions, we focus on the NEP and qNEP models.
To support reproducibility, all input files and trained DP models have been made publicly available via a GitHub repository~\cite{L75_data}.

\section{Results and discussion}

\subsection{AIMD Simulations of PbTiO$_3$}
PbTiO$_3$ is a prototypical and widely studied ferroelectric material. However, a rigorous benchmark of its finite-temperature behavior based on \textit{ab initio} methods is still lacking. Previous AIMD studies~\cite{Srinivasan03p168,Fang15p024104} were often limited by small simulation cells~\cite{Srinivasan03p168} or conducted in the canonical ($NVT$) ensemble with fixed lattice parameters~\cite{Fang15p024104}. The use of the $NVT$ ensemble is especially problematic, as it suppresses the large lattice fluctuations that are essential for capturing the ferroelectric-paraelectric phase transition. This constraint can lead to an overestimation of the predicted transition temperature, due to the strong coupling between strain and polarization in \PTO. 
To overcome these limitations and obtain statistically reliable results, we performed AIMD simulations in the $NPT$ ensemble using a $4\times4\times4$ supercell containing 320 atoms. In particular, simulations near the transition temperature were run for over 50~ps to ensure adequate sampling. To the best of our knowledge, this represents the largest AIMD study of \PTO~to date in terms of both system size and simulation duration.

\begin{figure*}
\centering
\includegraphics[width=1\textwidth]{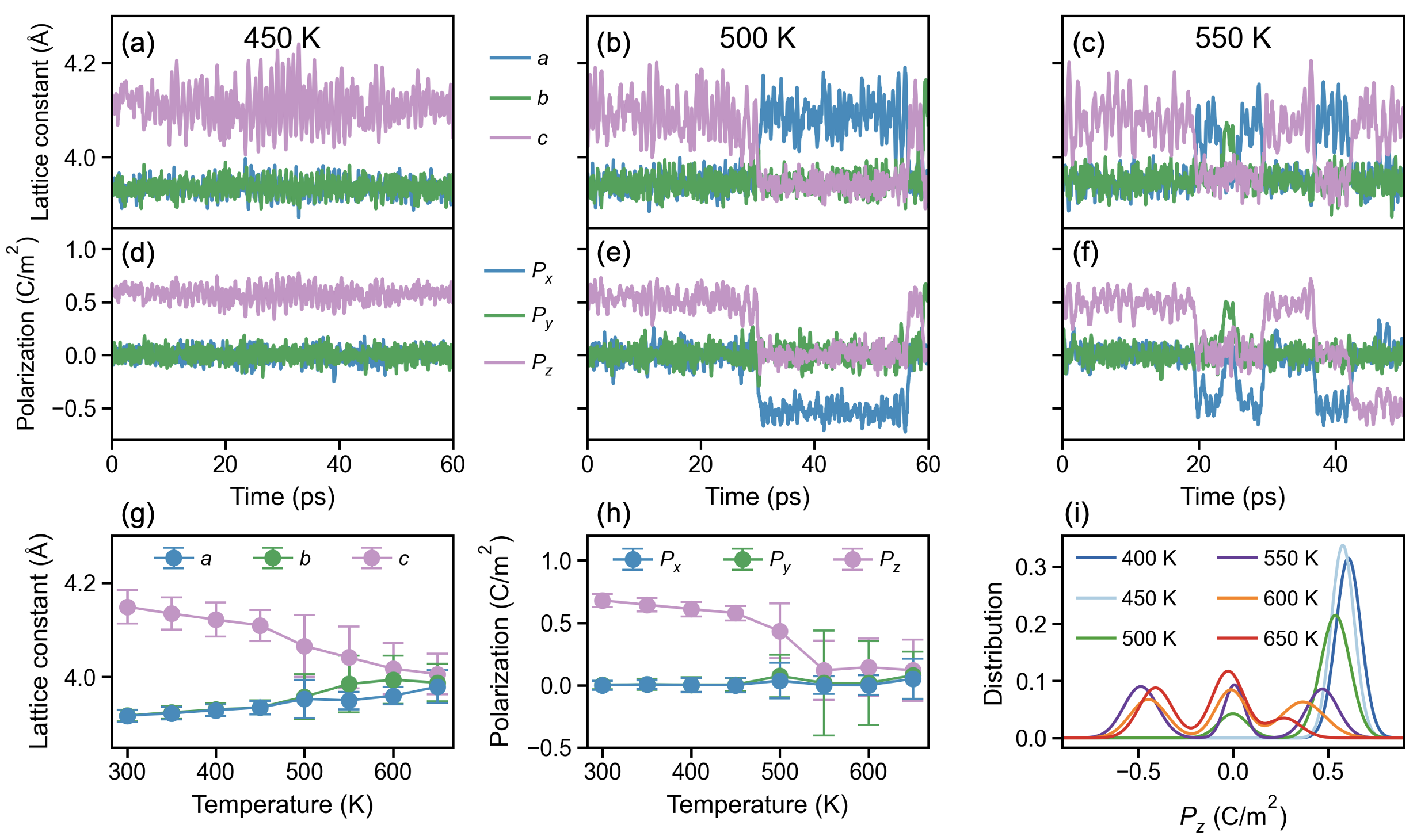}
\caption{
Time evolution of lattice constants and polarization fluctuations near the Curie temperature ($T_c$), obtained from $NPT$ ab initio molecular dynamics simulations using a $4 \times 4 \times 4$ (320-atom) supercell. Panels (a)–(c) present the lattice constants at 450, 500, and 550 K, respectively, while panels (d)–(f) show the corresponding polarization components along the Cartesian axes. A switching of the polar axis is observed at 500 K, indicating a transition temperature ($T_c$) of approximately 500 K.
Temperature dependence of (g) lattice constants and (h) polarization components along the Cartesian axes. Error bars represent the standard deviation, with significantly larger fluctuations observed above $T_c\approx 500$~K. Panel (i) shows the distribution of the polarization along the polar axis ($P_z$) of the supercell at various temperatures. At $T_c \approx 500$ K, a substantial population emerges near $P_z = 0$.
The analysis at each temperature is based on configurations sampled from the final 20 ps of the equilibrium AIMD trajectories.
}
\label{fig:pto-dft}
\end{figure*}

Figure~\ref{fig:pto-dft} shows the time evolution of the unit-cell lattice constants ($a$, $b$, $c$) and the corresponding Cartesian polarization components ($P_x$, $P_y$, $P_z$) from AIMD simulations of a 320-atom supercell at three representative temperatures.
At 450~K (Figs.~\ref{fig:pto-dft}a and \ref{fig:pto-dft}d), the system remains clearly ferroelectric. The polarization stays aligned along the $c$-axis ($z$-axis) with a magnitude of approximately 0.56~C/m$^2$, accompanied by a noticeable tetragonal distortion ($c/a > 1$).
At 500~K (Figs.~\ref{fig:pto-dft}b and \ref{fig:pto-dft}e), the system enters a dynamic regime where the polarization direction spontaneously switches. Initially aligned along the $z$-axis, the polarization abruptly rotates to the $x$-axis after about 30~ps. This switching behavior indicates that the free energy barrier for a 90\degree~polarization rotation becomes comparable to the thermal energy ($k_BT$), allowing spontaneous reorientation of the polar axis. In finite supercells, polarization axis rotation is further promoted by the suppression of domain formation. Notably, this rotation is absent in larger systems, such as $10\times10\times10$ supercells (as discussed below). This behavior can be interpreted as a manifestation of the small free energy difference between the ferroelectric and paraelectric phases. In the thermodynamic limit, this near-degeneracy gives rise to local polarization fluctuations that average out to zero net polarization, resulting in a macroscopically cubic paraelectric phase. We thus identify the onset of this behavior as the transition point, placing $T_c$ near 500~K.
At 550~K (Figs.~\ref{fig:pto-dft}c and \ref{fig:pto-dft}f), the switching becomes more frequent. Throughout the simulations, the lattice constants and polarization components exhibit strongly correlated fluctuations, highlighting the significant electromechanical coupling in \PTO~and the importance of using variable-cell ($NPT$) simulations to capture the transition accurately. Even at this temperature, the instantaneous polarization along the long axis remains finite, reflecting the order–disorder nature of the transition. The paraelectric phase arises not from vanishing local dipoles, but from their dynamic averaging over time.

\subsection{Curie Temperature from AIMD}

The ensemble-averaged temperature dependence of lattice constants and polarization components (Figs.~\ref{fig:pto-dft}g-i) captures the characteristic ferroelectric–paraelectric phase transition in PbTiO$_3$. To ensure a consistent reference frame for analysis, we first averaged the atomic configurations over the full equilibrium trajectory at each temperature. The resulting average lattice constants were sorted as $a < b < c$, and the Cartesian coordinate system was defined such that the $c$-axis aligned with the $z$-direction. This fixed frame was then used to compute the instantaneous polarization components ($P_x$, $P_y$, $P_z$) throughout the trajectory.

As shown in Figs.~\ref{fig:pto-dft}g and h, the average length of the $c$-axis decreases with increasing temperature, accompanied by a gradual reduction in the ensemble-averaged $P_z$. The variances (indicated by error bars) increase with temperature, reflecting enhanced thermal fluctuations. At 400 and 450~K, the system exhibits low variance and sharp, unimodal $P_z$ distributions (Fig.~\ref{fig:pto-dft}i), consistent with a stable ferroelectric state.
Above 500~K, this behavior changes significantly. The emergence of a secondary peak at $P_z = 0$ leads to broader, multimodal distributions and larger error bars, suggesting the onset of polar instability. The gradual shift of peak positions toward lower polarization values with increasing temperature indicates a displacive softening of the polar mode. 

At 550~K and above, the distributions become distinctly trimodal, with two symmetric peaks at positive and negative values and a central peak at zero. This reflects dynamic switching between oppositely polarized states and a non-polar intermediate, resulting in reduced average $P_z$ at elevated temperatures. Taken together, these results indicate that the phase transition in PbTiO$_3$ exhibits features of both displacive and order–disorder mechanisms.

Consequently, although PBEsol overestimates the zero-temperature $c/a$ ratio (1.08 compared to the experimental value of 1.06~\cite{li25p21538}), our large-scale AIMD simulations using PBEsol and a $4\times4\times4$ 320-atom supercell yield a theoretical transition temperature of $T_c \approx 500$ K, lower than the experimental value of 760~K.

\subsection{Curie Temperature from DPMD}

\begin{figure*}
\centering
\includegraphics[width=0.7\textwidth]{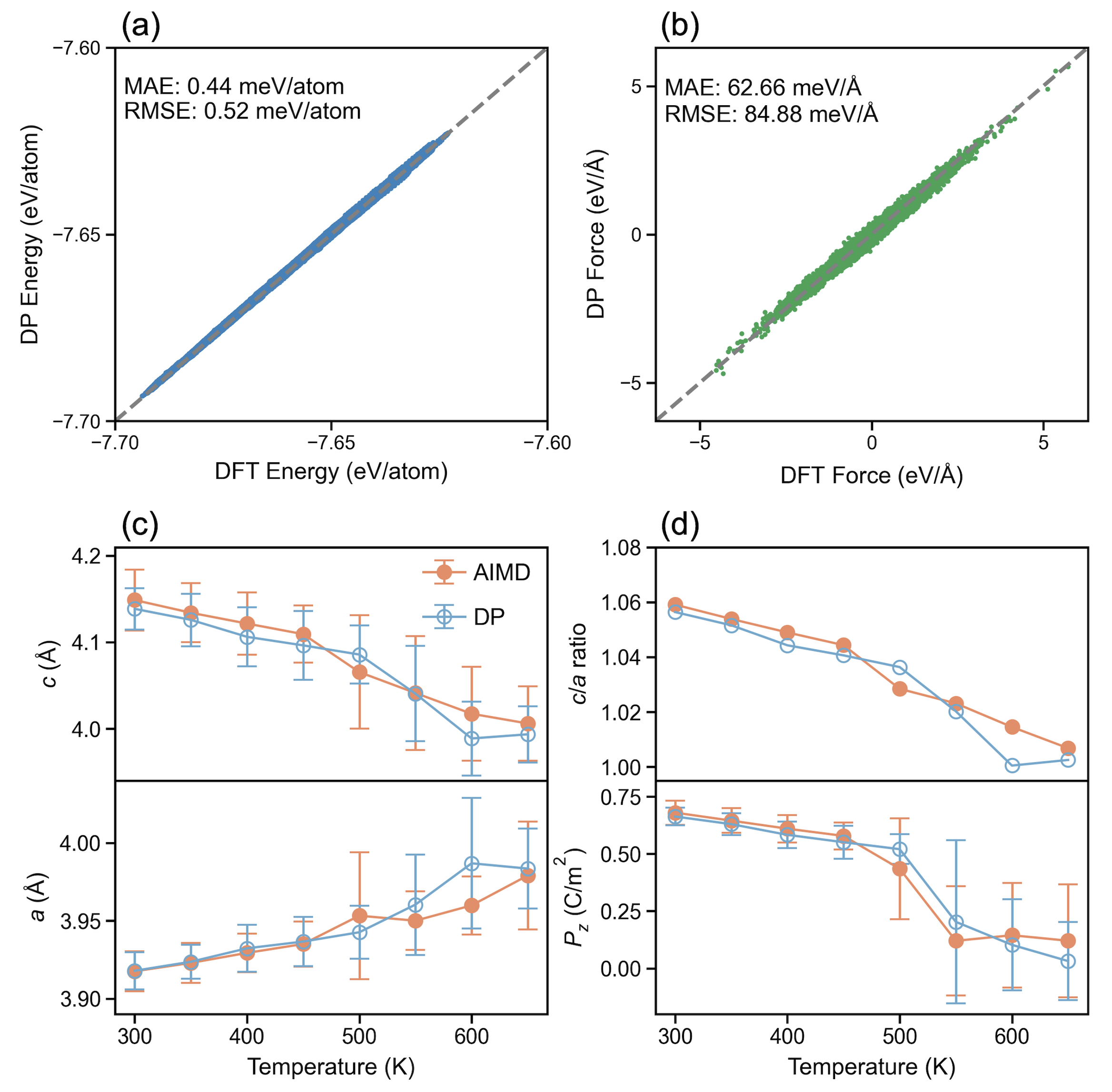}
\caption{
Comparison of DPMD and AIMD results.
Parity plots comparing (a) energies and (b) atomic forces predicted by the DP model against reference values from AIMD configurations not included in the training dataset. The fitting errors are comparable to those from the training set, indicating that the model is not overfitting and demonstrates good generalizability.
Panels (c) and (d) compare the temperature dependence of lattice constants, tetragonality, and polarization component $P_z$ as predicted by DPMD and AIMD. At each temperature, ensemble-averaged properties are computed using the same number of supercell configurations sampled over an equivalent duration of equilibrium trajectories.
}
\label{fig:pto-dp}
\end{figure*}

To enable a rigorous comparison with the AIMD results, we trained a DP model using a previously published, comprehensive dataset consisting of 13,021 diverse PbTiO$_3$ configurations (see Methods for details).
We evaluated the model performance on configurations sampled from $NPT$ AIMD trajectories used in the study of the ferroelectric–paraelectric phase transition (Figs.~\ref{fig:pto-dp}a–b).
The validation errors, 0.52 meV/atom for energies and 84.88~meV/\AA{} for forces, are comparable to those from the training set, indicating strong generalization to unseen configurations and minimal overfitting.

As shown in Fig.~\ref{fig:pto-dp}c, the temperature-dependent lattice constants predicted by the DP model agree reasonably well with the AIMD results. Based on the evolution of $P_z$ (Fig.~\ref{fig:pto-dp}d), the DP model also predicts a transition temperature of $T_c = 500$~K. These benchmarks confirm that the DP model accurately reproduces the \textit{ab initio} potential energy surface. The underestimation of $T_c$ is therefore attributed primarily to the intrinsic limitations of the PBEsol functional, rather than to fitting errors in the DP model.

\subsection{Data Sampling and Model Robustness}
The large number of configurations generated from our large-scale AIMD simulations also enables us to explore an interesting question: how does the diversity of training data affect model accuracy, especially in the context of phase transitions? To investigate this, we compared two distinct dataset construction strategies. The first follows the widely used DPGEN concurrent learning protocol, which is designed to sample the potential energy surface broadly by iteratively identifying and including high-uncertainty configurations~\cite{Wu21p024108}. The DP model employed in this work was trained using a dataset generated through this approach. In contrast, conventional AIMD trajectories sample configurations according to the underlying thermodynamic distribution and primarily sample configurations within the equilibrium basin. We directly compare the DPGEN-generated dataset with another constructed exclusively from configurations extracted from our $NPT$ AIMD simulations.

\begin{figure*}
\centering
\includegraphics[width=1\textwidth]{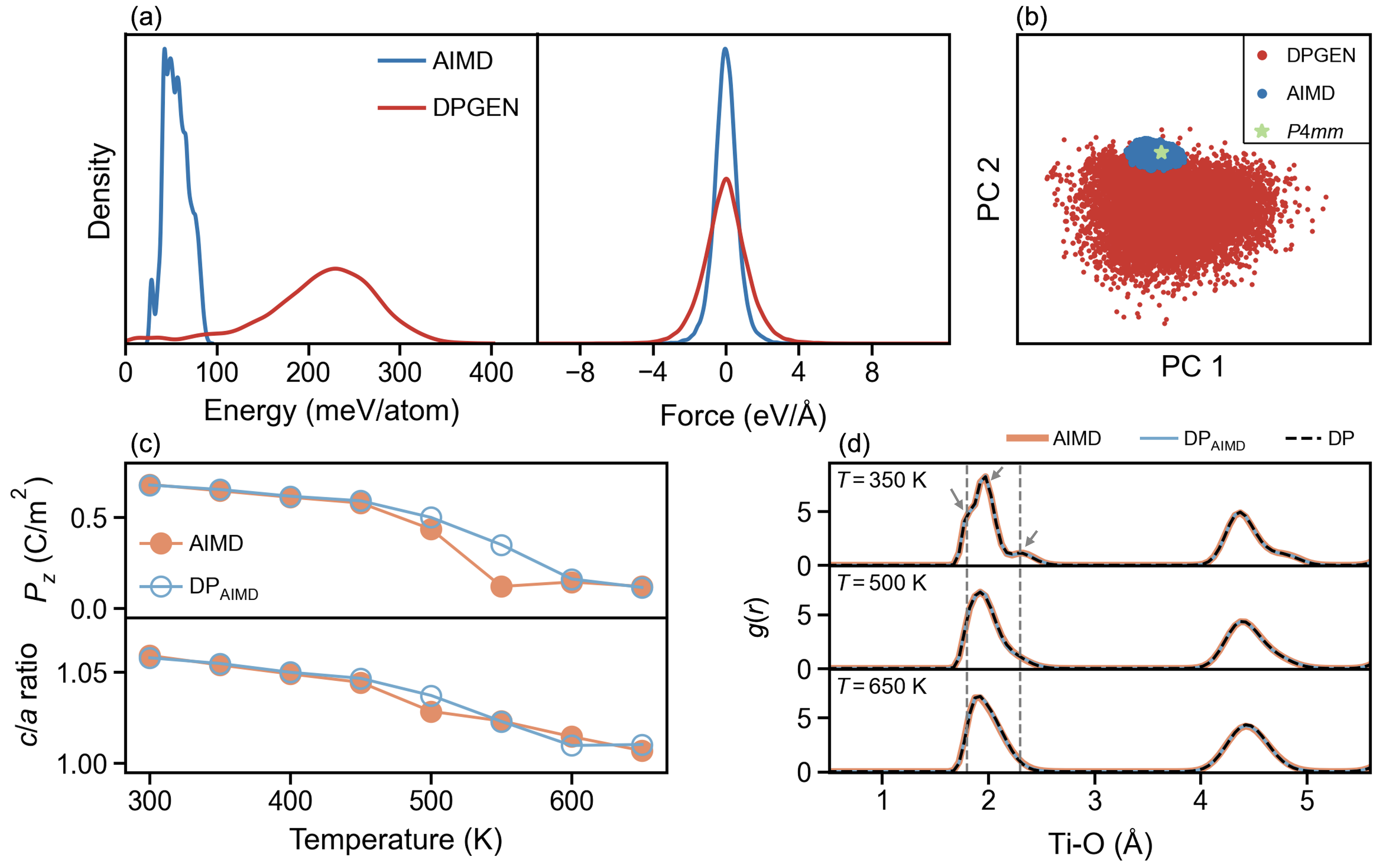}
\caption{
Comparison of configuration diversity between datasets from DPGEN and AIMD simulations.
(a) Distributions of energies (left) and atomic forces (right). The DPGEN dataset (red) spans a significantly broader range compared to the AIMD dataset (blue).
(b) Principal component analysis (PCA) of structural descriptors. The AIMD configurations (blue), obtained from a 310 ps trajectory, are confined to a narrow region near the tetragonal ($P4mm$) ground state. In contrast, the DPGEN dataset (red) explores a much broader region of configuration space. 
(c) Comparison of the temperature dependence of the polarization component $P_z$ (top panel) and the $c/a$ ratio (bottom panel), showing results from a DP model trained only on AIMD configurations (DP$_{\text{AIMD}}$) versus reference AIMD results.
(d) Comparison of radial distribution functions $g(r)$ obtained using different methods.
}
\label{fig:pto-data}
\end{figure*}

As presented in Fig.~\ref{fig:pto-data}a, the DPGEN dataset (red) spans a significantly broader range of energies and atomic forces compared to the AIMD dataset (blue). This contrast is further highlighted through Principal Component Analysis (PCA) of structural descriptors extracted from the DP model (Fig.~\ref{fig:pto-data}b). The AIMD configurations form a dense cluster around the tetragonal ($P4mm$) ground state, occupying a relatively narrow region of configuration space. In contrast, the DPGEN dataset forms a more diffuse distribution, reflecting a wider sampling of interatomic interactions.

To test whether this broader diversity is necessary for modeling the ferroelectric phase transition, we trained a separate DP model, denoted DP$_{\text{AIMD}}$, using only the configurations from the AIMD trajectory. Despite the narrower scope of the training data, the DP$_{\text{AIMD}}$ model accurately reproduces the temperature-dependent evolution of polarization ($P_z$) and tetragonality ($c/a$) as computed by AIMD (Fig.~\ref{fig:pto-data}c). Furthermore, the radial distribution functions $g(r)$ for Ti–O pairs at various temperatures, computed using AIMD, DP, and DP$_{\text{AIMD}}$, show nearly identical results (Fig.~\ref{fig:pto-data}d). 

These findings suggest that, for PbTiO$_3$, a dataset generated from sufficiently long (over 300 ps in total) $NPT$ AIMD trajectories near the transition temperature can provide adequate sampling to train a model that accurately captures the key characteristics of the ferroelectric phase transition. In this case, the trajectory naturally includes the relevant local structural variations driven by thermal fluctuations. However, this may not apply universally: the adequacy of AIMD-based datasets likely depends on the system and the complexity of its phase behavior. Therefore, we recommend a cautious approach when relying solely on AIMD trajectories for developing MLFFs.

\subsection{Finite Size Effects}

\begin{figure*}
\centering
\includegraphics[width=1\textwidth]{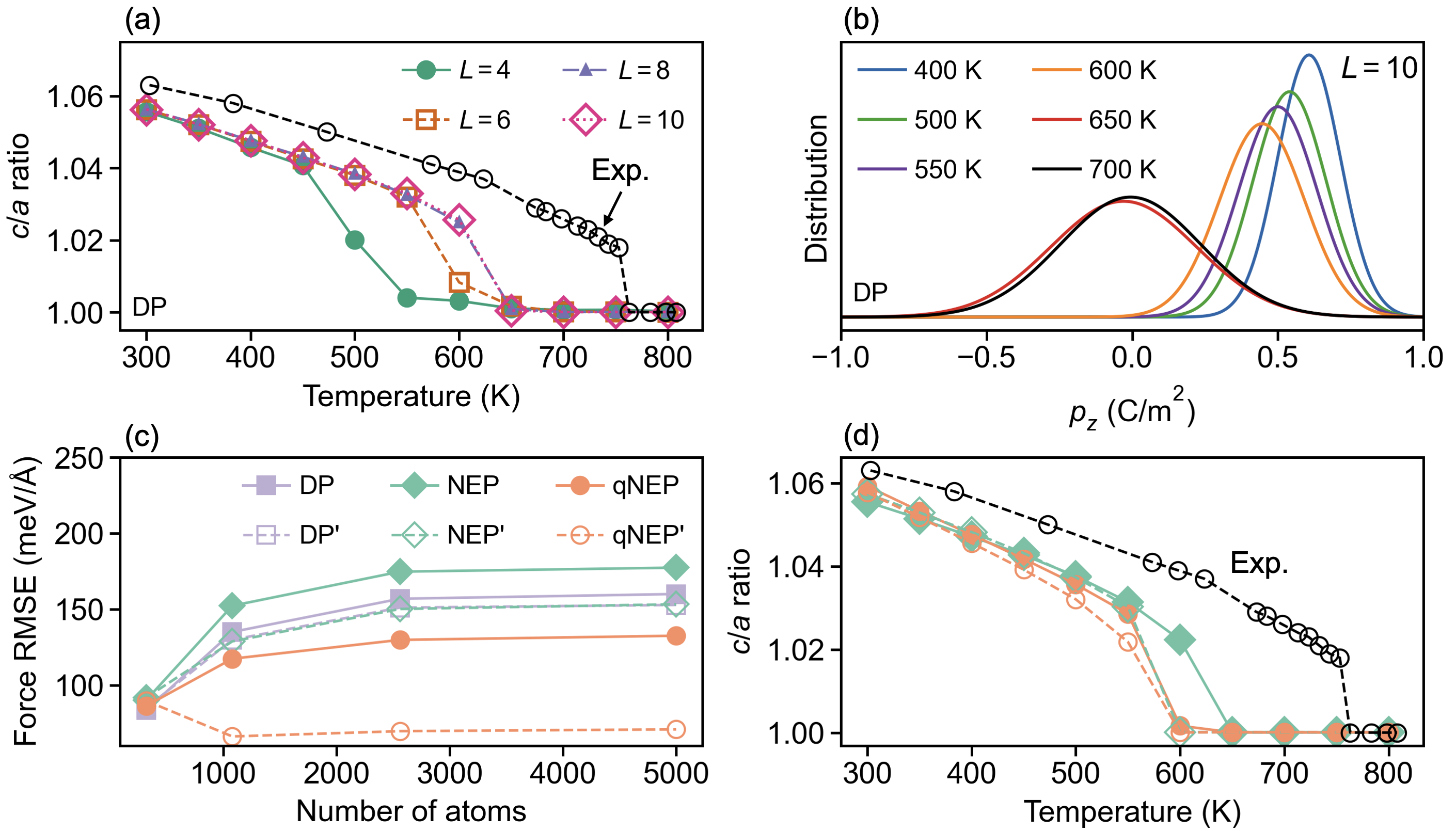}
\caption{
(a) Temperature dependence of the $c/a$ ratio for supercells of size $L \times L \times L$. The Curie temperature ($T_c$) converges at $L = 8$.
(b) Temperature evolution of the distribution of unit-cell-resolved local polarization components $p_z$, obtained from DPMD simulations using a $10 \times 10 \times 10$ supercell.
(c) Force root mean square error (RMSE) as a function of system size, relative to reference DFT calculations.
DP, NEP and qNEP models were trained exclusively on small-supercell (320-atom) configurations, while DP$^\prime$, NEP$^\prime$ and qNEP$^\prime$ models included large-supercell configurations in their training datasets. The short-range NEP and DP models exhibit comparable accuracy as well as systematic error accumulation with increasing system size, highlighting their limited scalability. In contrast, the long-range qNEP models maintain higher accuracy across system sizes. Incorporating large-supercell configurations has a marginal effect on the accuracy of short-range DP$^\prime$ and NEP$^\prime$ models but significantly improves the performance of qNEP$^\prime$.
(d) Temperature dependence of the tetragonal $c/a$ ratio obtained from $NPT$ molecular dynamics simulations using a large $12 \times 12 \times 12$ supercell. 
The local NEP models predict a higher Curie temperature, $T_c \approx 650$ K, than the long-range qNEP models.
In contrast, the qNEP and qNEP$^\prime$ models incorporating long-range interactions yield a converged intrinsic $T_c \approx 600$ K for the PBEsol functional. \note{The experimental data (black dashed line) were extracted from~\cite{Mabud79p49}.}
}
\label{fig:pto-finite-size}
\end{figure*}

The preceding comparison demonstrates that the DP model achieves DFT-level accuracy with minimal fitting errors for \PTO. Consequently, we can leverage this computational efficiency to explore finite-size effects on $T_c$ while maintaining the fidelity of first-principles calculations. 
A key advantage of DP is that its computational cost scales linearly with system size, which allows the simulations of large supercells that are otherwise computationally prohibitive for direct DFT calculations~\cite{Zhang18p143001}. This capability is particularly relevant for studying ferroelectric phase transitions, where long-wavelength phonon modes and domain wall dynamics could be strongly influenced by finite-size effects. To assess the convergence of $T_c$ with respect to system size, we performed simulations using the DP model with supercell sizes ranging from $4\times4\times4$ (320 atoms) to $10\times10\times10$ (5,000 atoms). As summarized in Fig.~\ref{fig:pto-finite-size}a, the predicted $T_c$ increases with system size and converges to approximately 650 K. \note{The black dashed line shows the experimental $c/a$ ratio, with a detailed comparison provided in Fig.~S6.} \note{A further dielectric susceptibility analysis with denser temperature sampling yields a consistent estimate of $T_c \approx 640$ K (see Fig.~S2).} \note{Multiple independent simulations with different initial states further confirm the robustness of the identified transition behavior near $T_c$ (see Fig.~S3).} This value is significantly closer to the experimental transition temperature of 760 K than the 450 K obtained from the smaller supercell, indicating that the limited spatial extent of the $4\times4\times4$ simulation artificially suppressed the phase transition. 

Additionally, DPMD simulations using a $10\times10\times10$ supercell did not exhibit the spontaneous rotation of the polar axis observed in AIMD simulations with a $4\times4\times4$ supercell. We computed the distributions of unit-cell-resolved local polarization $p_z$ at various temperatures, using 500 configurations sampled from a 50~ps equilibrium MD trajectory. The resulting $p_z$ distributions (Fig.~\ref{fig:pto-finite-size}b) reveal a gradual shift of the peak toward lower values with increasing temperature, characteristic of a displacive transition. At the critical temperature $T_c = 650$~K, the distribution becomes a broad Gaussian centered at zero. Notably, a significant number of unit cells still exhibit non-zero local polarizations, indicating that the macroscopic paraelectric phase emerges as an average over locally polarized regions, consistent with the general picture revealed by AIMD simulations (albeit using a smaller supercell).

\subsection{Long-range corrections}

The DP model depends on local descriptors to capture the local atomic environment. While computationally efficient, this local approximation inherently neglects explicit long-range electrostatic interactions, which could be important for accurately describing the properties of ferroelectric materials. A critical question thus arises: can a local model trained exclusively on small supercells (\textit{e.g.}, $N = 320$ atoms) retain its predictive fidelity when extrapolated to large-scale simulations ($N \ge 5000$ atoms) where long-range interactions become significant?

Unfortunately, the DP model lacks a long-range variant based on atomic point charges that could leverage the Ewald summation algorithm available in many MD packages. 
To investigate the effects of long-range interactions, we conducted a comparative study using both the standard NEP framework and its long-range extension, qNEP, which enhances the short-range descriptor with an electrostatics term derived from environment-dependent partial charges. Baseline models (DP, NEP and qNEP) were trained on the DPGEN dataset consisting of 320-atom supercells, computed using CP2K's OT method at the $\Gamma$-point. We then assessed their force prediction accuracy on progressively larger supercells, ranging from 320 to 5000 atoms.

As illustrated in Fig.~\ref{fig:pto-finite-size}c, the standard local models (DP and NEP) exhibit distinct error accumulation; the force RMSE nearly doubles as the system size increases from 320 to 1080 atoms and plateaus at a high error level. This degradation occurs because local descriptors fail to capture the variations in long-range electrostatic forces that emerge in larger supercells. In contrast, while the qNEP model also shows some error increase, it maintains significantly better accuracy than the standard NEP for larger supercells, demonstrating the effectiveness of explicitly including long-range Coulomb interactions.

To further investigate this size-dependency, we expanded the training dataset by including configurations from larger supercells and trained additional models, denoted as DP$^\prime$, NEP$^\prime$ and qNEP$^\prime$. Even with this enriched training set, the short-range DP$^\prime$ and NEP$^\prime$ models continue to exhibit considerable error for large supercells, confirming the fundamental limitation of the local approximation. Remarkably, the qNEP$^\prime$ model demonstrates superior transferability, maintaining a consistently low error baseline across all system sizes. This confirms that incorporating explicit electrostatics is essential for achieving size-independent accuracy in ferroelectric simulations. We note that qNEP$^\prime$ achieves even lower errors for large supercells than for the 320-atom case. This suggests that the explicit electrostatic term effectively captures the physics of the bulk limit, whereas the smallest systems contain complex finite-size artifacts and periodic image interactions that are inherently more difficult to fit.

While the superior accuracy of qNEP and qNEP$^\prime$ over their short-range counterparts might suggest that they would predict a transition temperature closer to the experimental value, the results reveal a more nuanced picture. Fig.~\ref{fig:pto-finite-size}d shows the temperature dependence of lattice tetragonality ($c/a$), computed using a large $12\times12\times12$ supercell. The standard NEP model (solid green line) predicts a transition temperature of $T_c \approx 650$~K, identical to previous DPMD results, indicating that both short-range models perform comparably. However, a distinct shift occurs when long-range interactions are included: both the qNEP and qNEP$^\prime$ models (orange lines) converge to a notably lower transition temperature of $T_c \approx 600$~K.

This systematic reduction suggests that the higher $T_c$ predicted by short-range models is likely a fortunate outcome, potentially caused by an artificial stiffening of the potential energy surface due to the truncation of electrostatic interactions. Consequently, it is reasonable to conclude that 600~K represents the true thermodynamic limit of the PBEsol functional for PbTiO$_3$ when long-range electrostatics are properly accounted for. The seemingly accurate $T_c$ predicted by short-range models appears to result from a fortuitous cancellation of errors.

\section{Conclusion}
In this work, we have systematically disentangled the origins of the persistent discrepancy between theoretical and experimental Curie temperatures $T_c$ in ferroelectrics, using PbTiO$_3$ as a prototypical case study. By conducting the largest constant-pressure AIMD simulations of PbTiO$_3$ reported to date and cross-validating them with machine learning force fields, we have isolated the specific sources of error inherent in first-principles phase transition modeling.

Our results conclusively demonstrate that the underestimation of $T_c$ in PbTiO$_3$ is not due to the machine learning fitting process. When trained on adequate data, the deep potential model reproduces the \textit{ab initio} potential energy surface with high fidelity, yielding a transition temperature identical to that of direct AIMD ($T_c \approx 500$ K). Consequently, the substantial deviation from the experimental value of 760~K is primarily attributable to the intrinsic limitations of the PBEsol exchange-correlation functional.
Furthermore, we have elucidated a critical interplay between finite-size effects and the locality of interatomic interactions. While increasing the supercell size generally favors a higher $T_c$ by accommodating long-wavelength fluctuations, we find that standard, purely local machine learning descriptors artificially stiffen the lattice. This leads to an increase of the transition temperature to 650~K in large systems. By employing the long-range qNEP formalism to explicitly account for long-range electrostatic interactions, we determined the true converged thermodynamic limit for PBEsol to be approximately 600 K.

This finding carries important implications: it suggests that improved agreement with experiment does not necessarily indicate a genuine improvement in model accuracy. Specifically, in the prediction of transition temperatures, the opposing effects of finite-size suppression and artificial stiffening caused by the neglect of long-range interactions can coincidentally lead to better agreement with experimental values. We conclude that accurate $T_c$ prediction requires not only high-quality training data and large simulation cells but also the explicit treatment of long-range electrostatics and, ultimately, the adoption of exchange-correlation functionals beyond the generalized gradient approximation.

\backmatter

\bmhead{Acknowledgements}

We acknowledge the supports from National Natural Science Foundation of China (92370104). The computational resource is provided by Westlake HPC Center. 

\section*{Declarations}

\begin{itemize}
    \item \textbf{Funding}
    This work was supported by the National Natural Science Foundation of China (92370104).

    \item \textbf{Conflict of interest}
    On behalf of all authors, the corresponding author states that there is no conflict of interest.

    \item \textbf{Data availability}
    The data that support the findings of this study are available in the~\cite{L75_data}.

    \item \textbf{Author contribution}
    S.L. and C.S.A. contributed to the study conception and design. D.N.L. performed the calculations. Data analysis was performed by D.N.L. and S.L. The manuscript was written by D.N.L., S.L., and C.S.A. All authors read and approved the final manuscript.
\end{itemize}

\bibliography{SL.bib}


\clearpage
\newpage

\clearpage
\newpage

\clearpage
\newpage

\clearpage
\newpage

\end{document}